\documentclass[aps,prx,twocolumn,amsmath,amssymb,superscriptaddress,preprintnumbers]{revtex4-1}
\usepackage{color}
\usepackage{graphicx}
\usepackage[normalem]{ulem}
\usepackage{braket}

\begin{document}

\title{Control Electronics For Semiconductor Spin Qubits}

\author{Lotte Geck}
\email[]{l.geck@fz-juelich.de}
\affiliation{Central Institute ZEA-2, Electronic Systems, Forschungszentrum J\"ulich GmbH, Germany }

\author{Andre Kruth}
\affiliation{Central Institute ZEA-2, Electronic Systems, Forschungszentrum J\"ulich GmbH, Germany }

\author{Hendrik Bluhm}
\affiliation{JARA-FIT Institute Quantum Information, Forschungszentrum J\"ulich GmbH and RWTH Aachen University, 52074 Aachen, Germany}

\author{Stefan van Waasen}
\affiliation{Faculty of Engineering, Communication Systems, University of Duisburg-Essen, Germany }
\affiliation{Central Institute ZEA-2, Electronic Systems, Forschungszentrum J\"ulich GmbH, Germany }

\author{Stefan Heinen}
\affiliation{Chair of Integrated Analog Circuits and RF Systems, RWTH Aachen University, Germany }

\preprint{preprint v1}

\date{\today}

\begin{abstract}
Future universal quantum computers solving problems of practical relevance are expected to require at least $10^6$ qubits, which is a massive scale-up from the present numbers of less than 50 qubits operated together. Out of the different types of qubits, solid state qubits are considered to be viable candidates for this scale-up, but interfacing to and controlling such a large number of qubits is a complex challenge that has not been solved yet. One possibility to address this challenge is to use qubit control circuits located close to the qubits at cryogenic temperatures. In this work we evaluate the feasibility of this idea, taking as a reference the physical requirements of a two-electron spin qubit and the specifications of a standard 65\,nm complementary metal-oxide-semiconductor (CMOS) process. Using principles and flows from electrical systems engineering we provide realistic estimates of the footprint and of the power consumption of a complete control-circuit architecture. Our results show that with further research it is possible to provide scalable electrical control in the vicinity of the qubit, with our concept.

\end{abstract}


\maketitle

\section{Introduction\label{sec:intro}}

Quantum computing promises exponential speedup and memory
enhancement for certain computational problems. However, applications such as Shor's algorithm 
will possibly require many
millions of qubits for problem sizes where an advantage over
conventional computers can be expected \cite{JonesMeterFowlerEtAl2012}. This large number mainly results from the need of making quantum systems robust against errors due to decoherence, i.e. against the loss of the information stored in the qubits caused by the interaction with the environment.  
The (theoretically) most established  approach to counteract these errors relies on the redundant encoding of the fragile quantum
information to enable error correction, resulting in an overhead of
$10^2$ to $10^5$ in the number of physical qubits.  Building on fabrication techniques that enable chips with billions of
transistors, solid-state qubit implementations are promising candidates to support scaling to such large numbers. However, connecting and controlling such a large number of qubits represents a formidable engineering challenge.

Current experiments on the most scalable solid state qubits
and processor demonstrators \cite{Kelly2015,Reagor2018}
are operated at temperatures of 20-100\,mK, and
manipulated using room temperature signal generators, whose output is routed to the qubits via coaxial cables.
As argued in Ref.\,\cite{DATE_2017},
this approach will become very
cumbersome beyond a few hundred qubits at the latest, and seems
completely impractical for the large number of qubits eventually required. Even with a fairly compact PCI-type form factor for a single qubit
controller, the control electronics would fill thousands of racks. Furthermore,
it poses a rather fundamental connectivity challenge. For example,
$10^6$ UT34 coaxial cables with a diameter of 0.9\,mm$^2$ each
correspond to a total cross section of 0.6\,m$^2$. Such cabling will
impose an unacceptable heat load on the cryogenic end, and cannot be
connected directly to the highly-integrated qubit chips -- current
semiconductor technology is limited to die sizes of about 30\,mm.
Multiplexing can help, but will likely be limited to a relatively
small number of signals per cable, due to frequency crowding and
throughput considerations.

\begin{figure}  
	\includegraphics[width=0.7\linewidth]{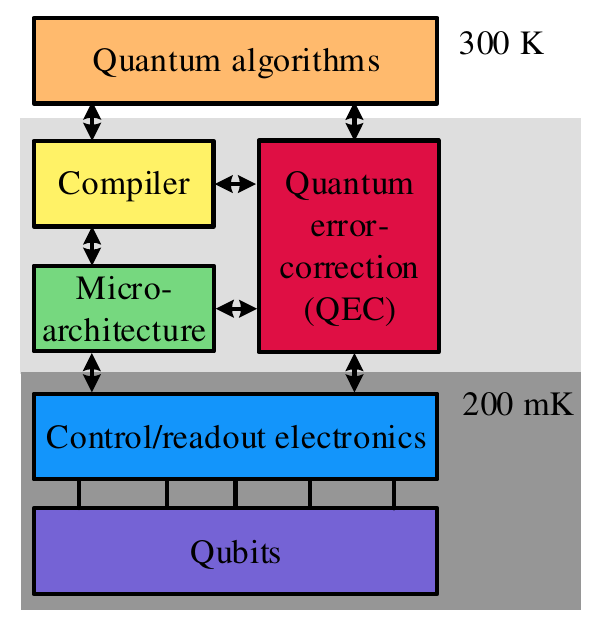}
	\caption{General architecture of a quantum computer, going from a high-level description of a quantum algorithm, to the actual physical operation of the qubits in a layered architecture \cite{Fu2016}.}\label{pic:levels}
\end{figure}

A proposed control solution is to rely on manipulating many
qubits with the same signal \cite{VeldhorstEeninkYangEtAl2016, VandersypenBluhmClarkeEtAl2017}, possibly
after selecting desired pulses and using modulators to tune the
amplitude or phase of a pulse \cite{HornibrookCollessLambEtAl2015}, or
to switch the qubit resonance frequency. However, this approach requires that the
qubits are very similar to each other (at least with respect to the
drive frequency), which has not been demonstrated so far for solid-state qubits. 
Furthermore, it does not address the need of gate-defined spin-qubits in semiconductors, where several DC voltages at the corresponding gates have to be tuned, 
which is difficult to achieve with one bias line.

All the above challenges could be addressed with highly integrated
control systems providing all required functionality in the
immediate vicinity of the qubits, either on the same chip or with some
form of chip-to-chip interconnection technology. The key advantage of this approach is that microfabricated interconnects could be used between the qubit and the control system.
This idea is consistent with the general vision of a layered architecture for quantum computer 
 sketched in Fig.\,\ref{pic:levels} \cite{Fu2016}. In this scheme, quantum algorithms are described by a high-level programming language that is agnostic to the faulty quantum hardware and assumes both idealized qubits and operations. These algorithms are converted into their fault-tolerant version based on a specific quantum error-correction-code in a lower level of the quantum computer, and converted into a series of instructions that include both the execution of the algorithm, and the detection and corrections of errors. The instructions are translated into a series of control waveforms to be applied to the qubit by the control electronics, which should be located as close as possible to the actual qubits. The focus of this work is on this lower layer.

Placing the qubit-control electronics at low temperature poses stringent constraints in terms of area and
power consumption. Ideally, each qubit should be associated with a control unit whose footprint corresponds to the average interqubit spacing.
Furthermore, the limited cooling power available at low temperature  is a major challenge for this approach. 
Determining the requirements for such a
dedicated controller is therefore a necessity for weighing the advantages
and disadvantages of different options for controlling large numbers of qubits.

Early works on cryoelectronic circuits relevant for qubit control
have largely focused on specific functionalities such as pulse generators \cite{EkanayakeLehmannDzurakEtAl2008}, readout amplifiers \cite{CurryEnglandBishopEtAl2015,TracyLuhmanCarrEtAl2016} and digital-to-analog
converters (DACs) \cite{RahmanLehmann2014}. In addition to these custom-designed integrated
circuits, complete field-programmable gate arrays (FPGAs) were shown
to remain largely operational at 4\,K, and used for flexible
implementations of key functionalities \cite{HomulleVisserPatraEtAl2016,SebastianoHomullePatraEtAl2017}.
These approaches show that modern
complementary metal-oxides-semiconductor (CMOS) technology can operate
at low temperature, even if the resulting power consumption is currently too large
for controlling a very large number of qubits below 1\,K.
Recent studies have made first steps to determine how the characteristics of individual electronics components change at
low temperature \cite{CharbonSebastianoVladimirescuEtAl2016,PatraIncandelaVanDijkEtAl2018}, which will be crucial for the design of optimized circuits.

Here, we present a well founded estimate of the area and power consumption of the control electronics for a single qubit. This estimate is based on the complete modeling of the control system, down to the level of elementary logic blocks (gates, flip-flops, etc.).  For concreteness, we designed the control system having in mind the requirements of two-electron spin-qubits \cite{Petta2005} (see Fig.\,\ref{pic:device}), but much of the insight
can also be transferred to other types of solid state qubits. Our control
system is able to translate digital commands from higher-level control-instances into analog control-waveforms applied to the qubit, and to supply stable DC-gate voltages as needed by semiconductor
qubits.  We
estimate that, when implemented in current 65\,nm technology, the circuit will
have a footprint of the order of $180 \times 180\,\mu\mathrm{m}^2$, and a power consumption of 190\,$\mu$W
per qubit. The latter comes mainly from the digital part of the control system, which could be possibly reduced up to four 
orders of magnitude by using a technology that is optimized for
low temperature operation. In addition new technology nodes could improve the area to $20 \times 20\,\mu\mathrm{m}^2$ and the power consumption to 700\,nW . These results indicate that highly
integrated control systems can be viable for at least $10^4$
qubits, and identify the advances required to realize them for systems
with various qubit numbers.

The remainder of this paper is organized as follows. In Sec.\,\ref{sec:requirements} we briefly summarize the operation of a two-electron-spin qubits
and the requirements it imposes on the associated electronic control system. In Sec.\,\ref{sec:modelling} we present 
our design for the architecture of the control system. 
The estimated area and power consumption of this system are presented in
Sec.\,\ref{sec:estimates}, and further discussed in Sec.\,\ref{sec:discussion}.  A summary and conclusion is given in Sec.\,\ref{sec:conclusion}.

\section{System requirements and target specifications\label{sec:requirements}}
\begin{figure}  
	\includegraphics[width=0.8\linewidth]{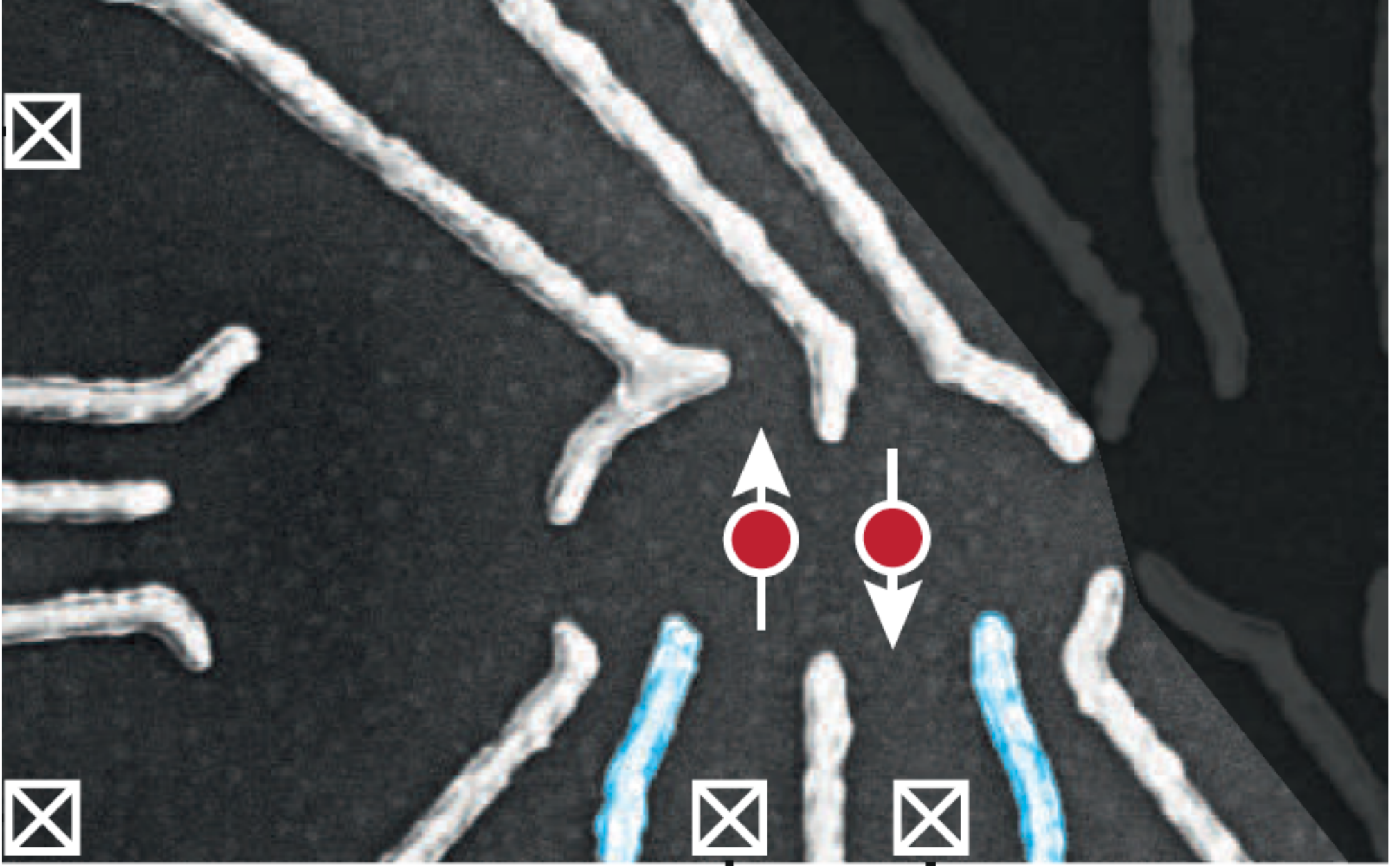}
	\caption{Typical device layout of a two-electron qubit \cite{Petta2005} in GaAs. The gray features correspond to the DC-gate electrodes used to form the two-electron-spin qubit in this work and the neighboring sensing dot (on the left). The  blue features are the RF-gate electrodes used for fast qubit manipulation.  The crossed boxes represent ohmic contacts
to the leads. }\label{pic:device}
\end{figure}

A two-electron spin-qubit is realized by confining two electrons into a double gate-defined quantum dot (see e.g. Fig.\,\ref{pic:device}). The latter is formed by exploiting the emergence of a two-dimensional electron gas (2DEG) at the interface between two semiconductors (e.g. GaAs/Al$_{x}$Ga$_{1-x}$As) or between a semiconductor and a dielectric (e.g. SiO$_2$), and by using metallic top-gates to deplete the 2DEG, creating gate-defined puddles of electrons (the quantum dots). The number of electrons in each dot can be controlled down to the single-electron regime via gate voltages. When occupied by two electrons, the double-dot can be operated as a singlet-triplet qubit, using the singlet $\ket{\textrm{S}}=\left(\ket{\uparrow\downarrow}-\ket{\downarrow\uparrow}\right)/\sqrt{2}$ and triplet $\ket{\textrm{T}_0}=\left(\ket{\uparrow\downarrow}+\ket{\downarrow\uparrow}\right)/\sqrt{2}$ as states  of the computational basis \cite{Burkard1999, Levy2002}, where the arrows $\uparrow$, $\downarrow$ describe the spin-state of the electron in each of the dots. The state of the qubit can be read out using a capacitively-coupled sensing dot, and exploiting Pauli blockade for spin-to-charge conversion \cite{Petta2005}. Qubit operations can be performed acting on the exchange coupling $J$ between the two electrons \cite{Levy2002} by means of electric pulses applied to two dedicated gate-electrodes \cite{FolettiBluhmMahaluEtAl2009}. An advantage of this control mode is that
it requires only baseband pulses (i.e. frequencies up to a few 100 MHz), eliminating the typical requirement of microwave control pulses of many other types of qubits.

\subsection{Target system specifications\label{subsec:spec}}
Common designs of two-electron qubits use up to six DC-gate electrodes for the qubit, and three for the sensing dot used for read-out (see Fig.\,\ref{pic:device}). We assume that one of these (e.g. a plunger gate for the sensing dot) can be eliminated, leading to a total of eight different DC signals that have to be managed by the control system ($N_{bias}=8$, see Tab.\,\ref{tab:spec}). These DC signals are typically in the volt range ($V_{range,bias}$=1\,V), while the resolution required to tune the qubit in the right operation regime is at the moment around 12\,bits. The stability of these signals, defined as root mean squared (RMS) voltage fluctuation, has to be below the level of quasi-static charge noise. We chose it to be 3\,$\mu$V
\footnote{\label{foot:1}The chosen value 3\,$\mu$\,V is below the level of quasi-static charge noise in current experiments, chosen so that improved device performance will not immediately lead to electronics-related performance issues and that the combined parasitic effects of the circuit still allow for enough signal quality.}.

The qubit is operated by applying control RF pulses to two dedicated electrodes. Each control pulse is formed by series of 16 different rectangular voltage pulses (samples). In current experiments, control pulses are produced by arbitrary waveform generators (AWGs), and have typically amplitudes of a few mV ($V_{range,RF}=4$\,mV), 10\,bit resolution and 1\,GHz sample rate, though a sufficient performance at 300\,MHz can be expected. The qubits frequency dependent sensitivity to broadband (white) noise is given through its filter function \cite{CywinskiLutchynNaveEtAl2007}, which shows a attenuating behaviour at higher frequencies \cite{CerfontaineBotzemDiVincenzoEtAl2014}. The noise spectral density must be at most $0.4\,$nV$/\sqrt{\text{Hz}}$ which translates to an RMS stability value of $\delta V_{RF}=8\,\mu$V.

Apart from generating the voltages to be applied to the DC and the RF gates, the control electronics must also be able to locally store the voltage configuration needed to define the qubit, as well as all the different control pulses used for  qubit operation. With surface-code error correction operation in mind, we consider up to $N_{pulses}=16$ different control operations, which includes initialization, readout, two-qubit gates with four neighboring qubits, and a few single-qubit gates. Some of the longer operations may have to be composed of several sequences. The requirements for the control electronics discussed above, 
and summarized in table \ref{tab:spec}, refer to parameters typical for GaAs-based devices, for which the control requirements have been
studied in more detail than for Si-based devices. In particular, our assumptions regarding noise, time resolution and control amplitudes are largely based on the detailed simulations of qubit-control pulses of Ref. \cite{CerfontaineBotzemDiVincenzoEtAl2014}, which uses a
realistic noise model based on the experiments of Refs.\,\cite{Dial2013,
	BluhmFolettiMahaluEtAl2010, BluhmFolettiNederEtAl2010} on GaAs-based devices. With these parameters 
fidelities of 99.8\,\% were predicted, and experimental results have
come close to these values \cite{CerfontaineBotzemHumpohlEtAl2016}. These numbers
exceed the error correction threshold, and come close to the target of
about 99.9\,\% at which the overhead for quantum error correction
becomes tolerable.  As Si-based devices are less
affected by dephasing caused by nuclear spins, they will probably allow
a lower operating speed than what is reported in Tab.\,\ref{tab:spec}, otherwise, the requirements will be very
similar.
\begin{table}
	\caption{Summary of current set of system specifications\label{tab:spec}}
	\begin{ruledtabular}	
		\begin{tabular}{l l l}
			 Specification& Symbol & Value \\
			\hline
			 Number of bias signals& $N_{bias}$ & 8\\
			 Bias range& $V_{range,bias}$ & 1\,V \\
			 Bias stability& $\delta V_{bias}$ & 3\,$\mu$V \\
			 Bias signal resolution& $n_{bias}$ & 12\,bit \\
			 Number of RF signals & $N_{RF}$ & 2 \\
			 RF amplitude & $V_{range,RF}$ & 4\,mV\\
			 RF signal resolution & $n_{RF}$ & 10\,bit\\
			 RF stability & $\delta V_{RF}$ & 8\,$\mu$V \\
			 RF sample rate& $f_{sample,RF}$ & 300\,MHz \\
			 RF pulse length& $l_{pulse}$ & 16\,samples \\
			 Number of RF pulses stored& $N_{pulses}$ & 16 \\
		\end{tabular}
	\end{ruledtabular}
\end{table}

A complete electrical system for qubits will also have to include a solution for
reading out qubits. As its performance requirements depend strongly on
the readout approach at the qubit level (e.g. charge sensor
\cite{Barthel2009} vs. gate-dispersive readout \cite{CollessMahoneyHornibrookEtAl2013}) and on the analogue
characteristics of the first amplifier stage, different kind of
considerations are required than for qubit manipulation. Hence, the focus of
the present work is entirely on the latter.

\begin{figure}  
\includegraphics[width=0.8\linewidth]{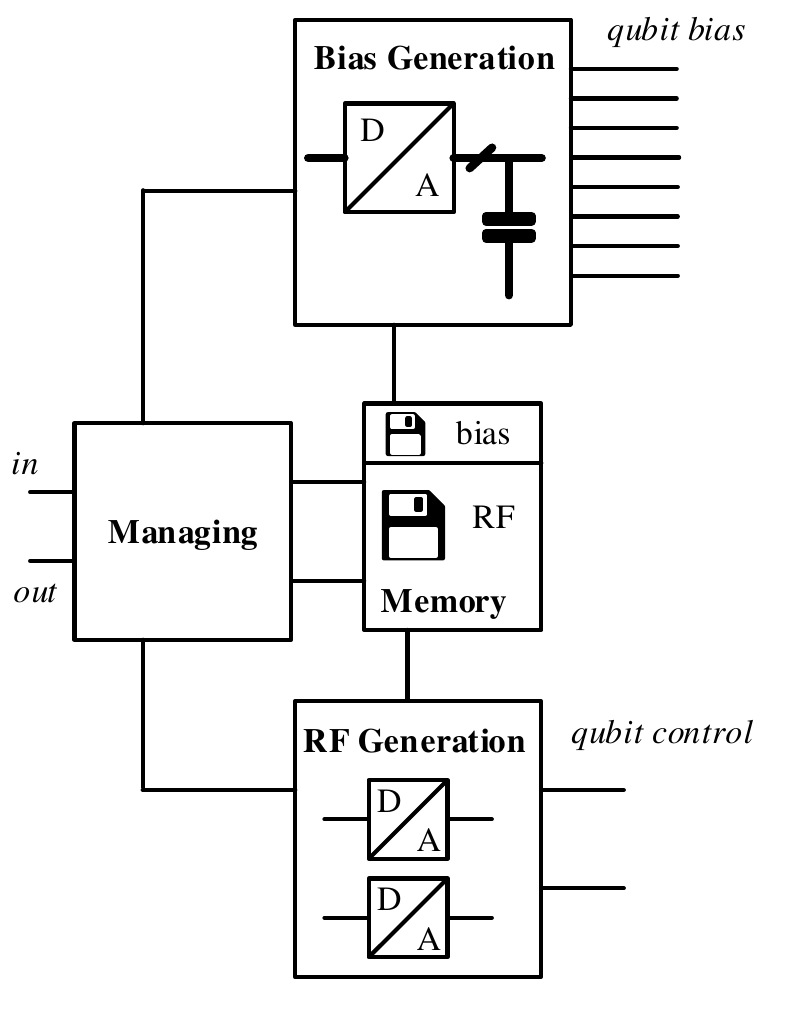}
\caption{Block diagram of the control electronics.  The managing unit steers the whole system and interacts with the higher levels of the quantum computer architecture. The bias memory stores values of the bias that needs to be applied to the DC electrodes for defining the singlet-triplet qubit. These values are converted into voltages by the bias generation unit, which supplies the DC electrodes of the double dot. The RF memory stores the control pulses needed to operate the qubits, which are converted into analogue signals applied to the control gates by the RF generation unit. }\label{pic:complete}
\end{figure}


\section{System design}  \label{sec:modelling}
The general architecture of the qubit-control electronics is sketched in 
Fig.\,\ref{pic:complete}. It consist of four units dedicated to specific tasks.
The managing component is a digital unit, which controls the
other subunits and interacts with the
higher levels of the quantum computer stack \cite{Fu2016}. The
memory stores the
voltage values to be applied to the DC-bias electrodes (bias memory) and the RF-pulse
sequences needed for qubit operations (RF memory). The bias generation unit
converts the digital values stored in the bias memory into voltages applied to the DC gate-electrodes, while the RF generation unit
generates the analog control pulses from the sequences stored in the RF memory.


\subsection{Memory\label{subsec:mem}}
\begin{figure*}
	\includegraphics[scale=0.72]{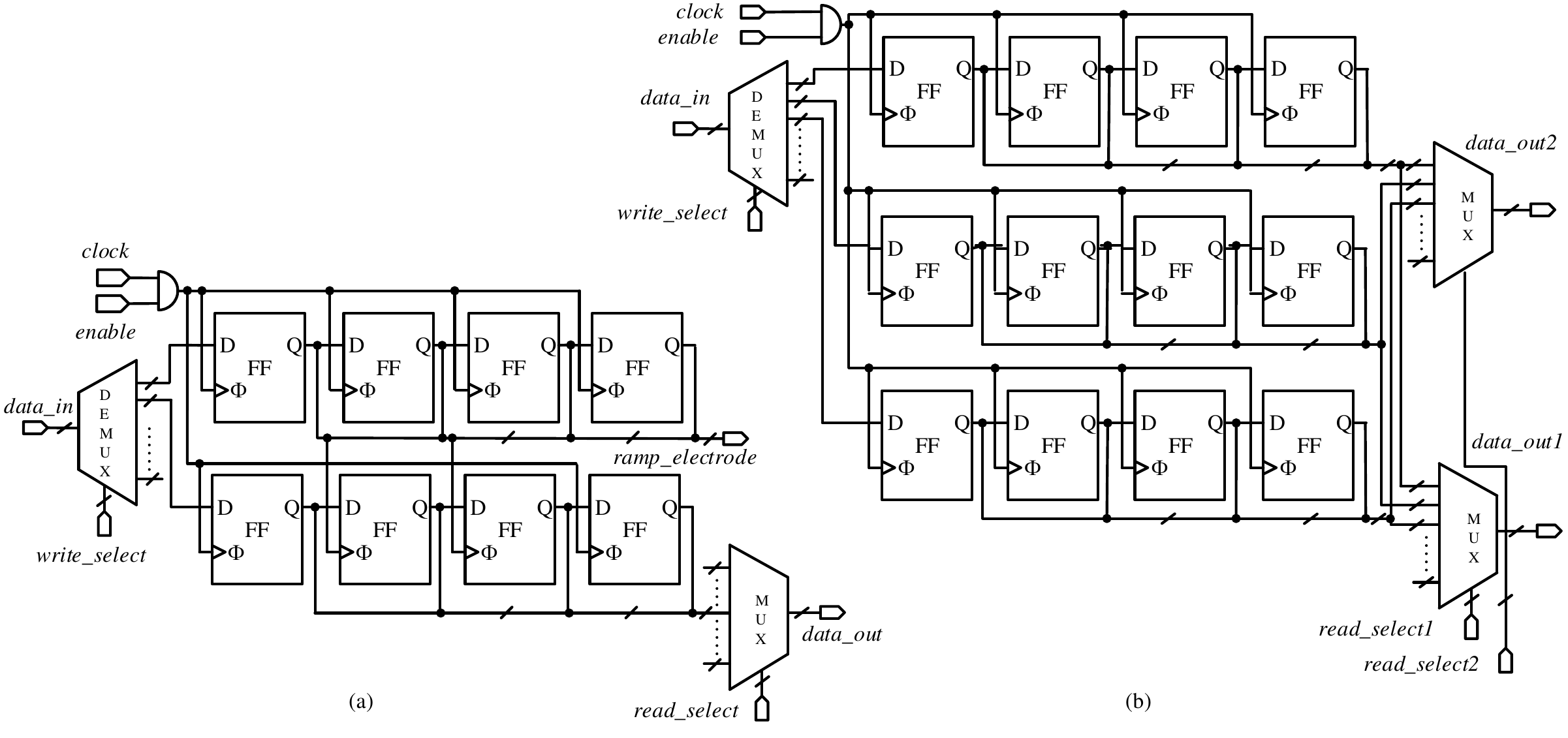}
	\caption{Structure of a) two registers of bias memory, and b) three register of RF memory for the case of 4\,bit resolution. In both cases, the registers have a serial write mechanism and a parallel readout. Read and write addressing is implemented through MUXs and DEMUXs. The output of the MUX {\em data\_out} connects the memory to the DACs inside the bias- and RF-generation part, while the select signals come from the managing unit. Symbols are defined in Fig.\,\ref{pic:managing-1}c. 
	\label{pic:memcomplete}}
\end{figure*}

The bias memory contains nine registers, one for each DC electrode ($N_{bias}=8$), plus an additional one that can be used to apply voltage ramps during the initial qubit-tune-up procedure \cite{Botzem2018}.
The RF memory requires considerably more space to store $N_{pulses}\cdot l_{pulse} = 256$ words.

For the memory, we have chosen a design based on flip-flops (FF) as they are more robust
with respect to variability in the component parameters than e.g. static
random access memory (SRAM) cells, which would be much more economical
in the number of transistors per bit. This conservative approach will
facilitate the fast implementation of a functioning demonstrator circuit
with currently commercially available technology. 

The resulting circuit design is shown in Fig. \ref{pic:memcomplete}. The use of flip-flops in a shift register configuration allows for a serial writing process with low wiring complexity. A serial write operation shifts a data-word into one register from the input {\em data\_in} in several clock cycles. The register is selected through a demultiplexer (DEMUX) with the corresponding address at {\em write\_select}.
The slow serial writing speed is not critical, because writing occurs only when a new set of pulse shapes/bias voltages is loaded into the memory outside of qubit operations.   

The read process on the other hand is more time critical,  
and it is implemented as a parallel readout in both memory parts, which allows reading all bits of a register in a single clock cycle. For this, the output of each flip-flop is connected to a multiplexer (MUX). The select signal {\em read\_select} determines the register whose values should be transmitted to the output {\em data\_out}. The main difference between the bias and the RF memory is that the second includes also a second MUX connected to all flip-flops, with an associated {\em read\_select} signal, which enables the parallel readout of two registers.


\subsection{Bias Generation\label{subsec:bias}}

\begin{figure}
	\includegraphics[scale=0.8]{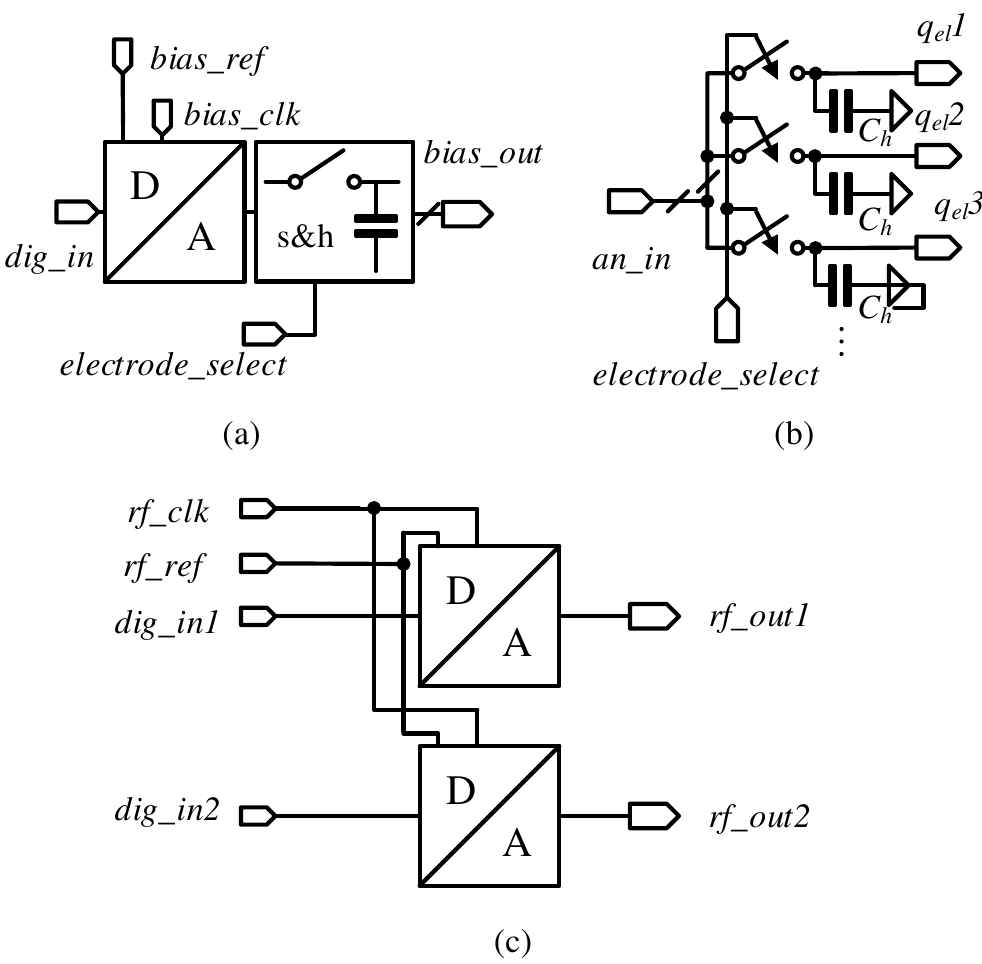}\vspace{0.4cm} 
	\caption{ (a) Structure of the bias generation unit. (b) Structure of the sample-and-hold for the bias generation unit. (c) Structure of the RF generation unit. Symbols are defined in Fig.\,\ref{pic:managing-1}c.
		 \label{pic:biasgen}}
\end{figure}
The bias generation unit is shown in Fig.\,\ref{pic:biasgen}a. It is composed of a DAC followed by a sample-and-hold (s\&h)
circuit, which allows using a single DAC to supply several bias electrodes.

\begin{figure*}
	\includegraphics[scale=0.8]{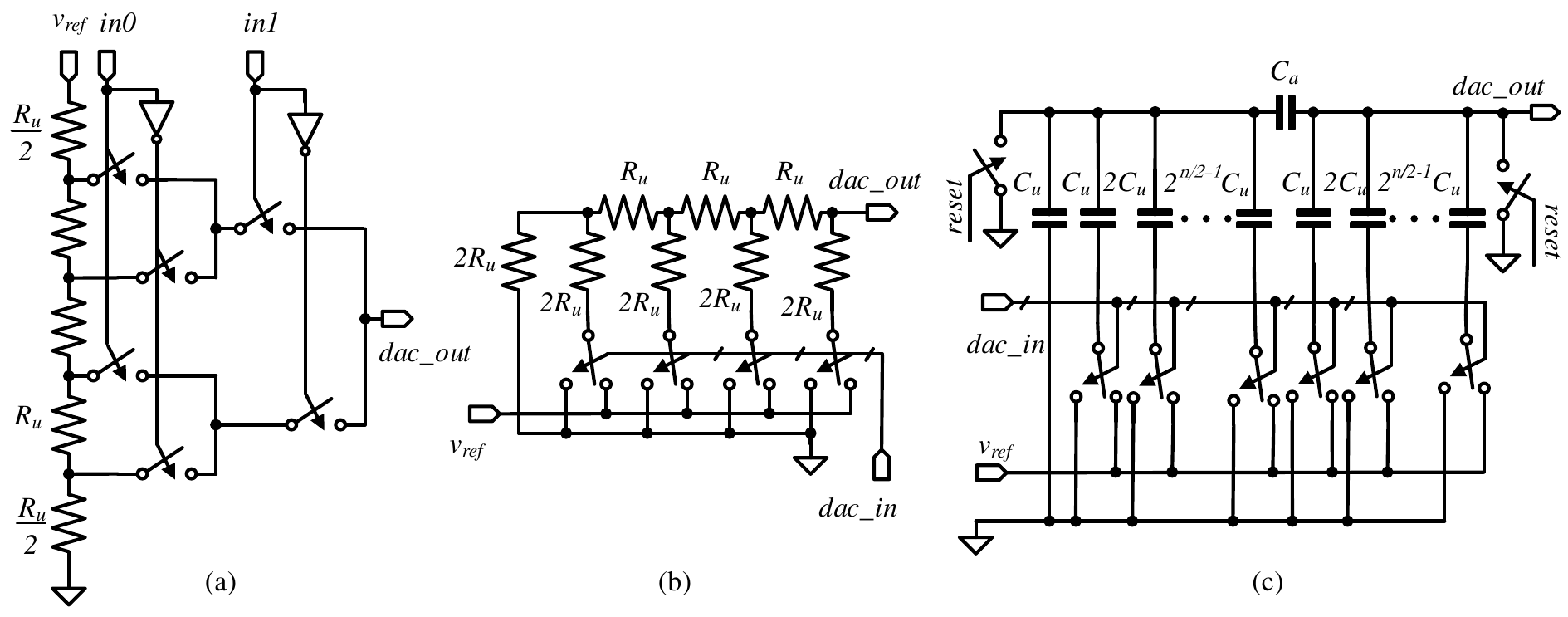}
	\caption{Basic DAC types\label{pic:kelvin} with reference voltages $v_{ref}$ or reference currents $i_{ref}$. $R_u$ denotes unit resistor, $C_u$ unit capacitor. (a) 2\,bit Kelvin Divider DAC consisting of one voltage divider string of resistors. b) 2\,bit R-2R Ladder DAC adds up fractions of the reference of the voltage or current at the output. (c) n-bit capacitor based DAC (Cap DAC). Here $C_a=C_{u}\cdot2^{n/2}/({2^{n/2}-1})$.   \label{pic:dac}}
\end{figure*}
Determining the ultimate best DAC architecture for the control electronics of a qubit is a complex task, which goes beyond the scope of this work.
There are several different ways of constructing a DAC \cite{Maloberti2007} and each is characterized by a number of properties such as range, stability, linearity, monotonicity, and noise, as well as by a characteristic area and power consumption. Here we consider only these last three most relevant aspects, and compare the three designs presented in Fig.\,\ref{pic:dac} (see discussion in Sec.\,\ref{sec:estimates}). These architectures are (i) the Kelvin Divider DAC (Fig.\,\ref{pic:kelvin}a), which uses a string of resistors as voltage dividers to tap different voltage levels from a voltage reference $v_{ref}$,  (ii) the R-2R Ladder DAC (Fig.\,\ref{pic:kelvin}b), which adds up fractions of the reference voltage at the output, and (iii) the Cap DAC (Fig.\,\ref{pic:kelvin}c), which works as a capacitive voltage divider between a reference voltage $v_{ref}$ and ground.  
The number of switches and resistors/capacitors needed for each DAC depends on the resolution $n$ as given in Tab.\,\ref{tab:dacs}.
We exclude from our analysis oversampling DACs, as they require significant additional circuitry and high clock frequency, as well as current-steering DACs, which are not very well suited for operations at cryogenic temperatures \cite{DasLehmann2010}.
Leveraging the very high qubit impedance -- which implies that next to no current needs to be delivered by the DACs -- we do not include operational amplifier as output buffer of the DACs. Errors caused by the output load cause no concern, since they are constant and calculable, and can 
be compensated by pre-distortion.

The size of the unit elements of the DACs, $R_u$ and $C_{u}$ are constrained by the requirement that the variations of the output voltage due to Johnson noise are within the stability criteria summarized in Tab.\,\ref{tab:spec}. For instance, the Johnson noise of the Cap-DAC, is approximately given by $\overline{v_{n}^{2}}=k_{B}T/C_{out}$, with $C_{out}\approx2^{n/2}C_{u}$. For a 12\,bit DC-bias DAC with $\delta V_{n}=3\,\mu$V and T=200\,mK, one obtains a lower bound of $C_{u}=4.8$\,fF. In the following, we assume for the Cap-DAC Metal-Insulator-Metal (MIM) capacitors with unit capacitance $C_{u}=10$\,fF, which is the minimal value allowed by the 65\,nm CMOS technology considered for our estimates (see discussion in Sec.\,\ref{sec:estimates}). The Johnson-Nyquist noise of the resistor based DACs is approximately given by $\overline{v_n^2}=4k_B T R_{out} B$, where $B$ is the effective bandwidth. For the Ladder DAC, it is $R_{out}=R_u$, assuming a bandwidth of 10\,MHz and T=200\,mK one obtains an upper bound of  $R_u\leq81$\,k$\Omega$. The output resistance of the Kelvin DAC is code-dependent and in the worst case scales $R_{out}=2^{n-2} R_u$ \cite{MarcheSavaria2010}. giving $R_u\leq160\, \Omega$ to remain within the noise requirements for a 12\,bit DAC and $R_u\leq5\, \Omega$ for 16\,bit. For the resistors forming the Kelvin DAC, we assume poly-silicon resistors with resistance $R_{u}=15\,\Omega$ for the starting comparison, which is again the minimal value set by the considered technology, keeping in mind that for $n=15,16$ this will violate the noise requirements. For the resistors of the Ladder DAC, we take instead the value $R_u=150\,\Omega$, as lower values of $R_u$ will lead to an excessive power dissipation (see discussion in Sec.\,\ref{sec:estimates}). In this context the bandwidth is associated with the overall low pass behaviour of the circuit. For this reason the bandwidth value during the design is engineered to let the wanted signal (1.1\,MHz) pass undistorted with margin (leading to B=10\,MHz), while unwanted signals and noise at higher frequencies are suppressed. Since the intrinsic DAC bandwidth is expected to be considerably higher than 10\,MHz, the resulting bandwidth could, for example, be lowered with additional capacitances connected in a low pass configuration.

The second element of the bias generation unit is a sample-and-hold circuit consisting of hold capacitors and CMOS
switches (see Fig.\,\ref{pic:biasgen}b). Each hold capacitor, $C_{h}$,
stores the voltage for a specific electrode, while the switches connect the DAC output to one capacitor at a time in order to
change and refresh its value. The refresh happens in a cyclic manner to compensate for
leakage-induced discharges. The s\&h circuit additionally
acts as a low-pass filter that decouples the qubit from high frequency
noise and glitches e.g. originating from switching in the DAC. Also in this case, 
the minimal value of $C_{h}$ is set by the requirement that the Johnson
noise of the s\&h, which is approximately given by $\overline{v^{2}}= k_{B}T/C_{out}$ with $C_{out}=N_{bias}C_h$, does not
exceed the specified voltage stability $\delta V_{bias}$. Assuming $T=200$\,mK for the electronics
and $\delta V_{bias}=3\,\mu$V, this gives $C_h\ge 38.4$\,fF. In the following we take the larger value $C_h=307\,$fF, which gives a better power vs area trade-off. The reason is that the value of $C_{h}$ contributes to the calculation of the rate $f_{refresh}$ at which the voltage on the capacitors have to
be refreshed to compensate for leakages. This sets the frequency at which all components associated with the bias generation work.  
For typical MIM capacitors  with a dielectric thickness of 10-30\,nm \cite{TSMC2004}  and a maximal applied voltage of 1\,V, the leakage current through the dielectric can be neglected with respect the leakage through the open switches with resistance $R_{off}$ (see Tab.\,\ref{tab:parameters}).
The refresh rate to meet the stability requirement is then given by
\begin{equation}
f_{refresh}=\frac{V_{range,bias}/R_{off}\cdot N_{bias}}{\delta V_{bias}\cdot C_{out}}.
\label{eq:refresh}
\end{equation}
where the numerator is the leakage current and the denominator the maximal charge that can be lost without compromising the stability of the voltage at the gate electrodes.
For given values, one obtains
$f_{refresh}=1.1$\,MHz.  This frequency determines also the operation frequency of the bias DAC.

\begin{table}
	\caption{DAC elements for resolution $n$ \label{tab:dacs}}
	\begin{ruledtabular}	
		\begin{tabular} {l c c c}
			& Kelvin & Ladder & Cap \\
			\hline
			Unit element & $R_u$ & $R_u$&$C_u$ ($C_a$)\\
			Unit value & 15\,$\Omega$ & 150\,$\Omega$ & 10\,fF\\
			No. unit elements & $2^n$ &$3n$&$2\cdot 2^{n/2}-1$\\
			No. switches & $2^{n+1}-2$ & $2n$& $2n$\\
		\end{tabular}
	\end{ruledtabular}
\end{table}

\subsection{RF Generation\label{subsec:rf}}

The RF generation unit is composed by two DACs, one for each RF electrode of the qubit as depicted in Fig\,\ref{pic:biasgen}\,(c).  This simple structure allows for fast waveform playback (300 MS/s, cf. table
\ref{tab:spec}). Similarly to what discussed in the previous section, the size of the unit components of the RF-DACs are determined by requiring that the voltage noise at DAC output is below the stability threshold. Using $n=10\,$bit, $B=600\,$MHz and $\delta V_{RF}=8\,\mu$V, this leads to the bounds $C_u\geq1.3\,$fF and $R_u\leq38\,\Omega$ (Kelvin) and $R_u\leq10$\,k$\Omega$ (Ladder). In the following, we will take for the unit elements of the RF DAC the same values chosen for the bias DAC and summarized in Tab.\,\ref{tab:parameters} for a first comparison, even if for $n\geq10$ the Kelvin DAC would produce too much noise. The bandwidth is set in accordance with the filtering behaviour of the qubit.

\subsection{Managing Component} 
The managing component is a purely digital unit which steers the other units of the control electronics, interacting with the higher levels of the quantum computer and providing an interface to the memory. 
Its structure is described in detail in Appendix~\ref{app:managing}.

\section{Area and power consumption}  \label{sec:estimates}

To make concrete estimate of the area and power consumption of the control system described in the previous section, we take as reference a standard commercial 65\,nm CMOS technology, which is well established and easily available for research purposes. The relevant parameters used  are summarized in Tab.\,\ref{tab:parameters}. Here, the resistive and capacitive densities presented in the first block are effective densities, which include device terminals. The second block  of parameters characterizes a typical transistor used as switches or in digital gates.
The third block of parameters contains the minimal resistance of poly-silicon resistors, the minimal capacitance of integrated MIM capacitors, and the digital supply voltage $V_{dd}=1\,$V set by process restrictions. All parameters of these first three blocks, are characteristics of the considered CMOS technology, and derived from the models in the physical design kit (pdk) \footnote{\label{foot:2} The pdk is provided by the manufacturer and subject to non disclosure agreements}. 
The fourth block contains our choices for the sizes of the resistors and capacitors inside the bias and RF generation parts. The last block of parameters describes the dynamic behaviour of the circuits in our model, including the switching rate of logic switches and additional clock frequencies. 
 
\begin{table}
	\caption{Parameters for 65\,nm CMOS technology used in the estimates of area and power consumption.
	\label{tab:parameters}}
	\begin{ruledtabular}	
		\begin{tabular}{l c c}
			Parameter & Symbol & Value \\
			\hline
			Resistive density & $\rho_R$&21.4\,$\Omega$/$\mu\text{m}^2$ \\
			Capacitive density &$\rho_C$ & 1.75\,fF/$\mu\text{m}^2$ \\
			\hline
			Mean transistor area & $A_{MOS}$& $0.375 \,\mu\text{m}^2$ \\
			Mean transistor cap. & $C_{MOS}$ & 150\,aF\\
			Mean transistor off res. & $R_{off}$ & 1\,T$\Omega$ \\
			Mean transistor on res. & $R_{on}$ & 5\,k$\Omega$\\
			\hline
			Min. Resistance & $R_{min}$ & $15\,\Omega$\\
			Min. Capacitance & $C_{min}$ & 10\,fF\\
		   Digital supply voltage& $V_{dd} $ & 1\,V\\

			\hline
			Hold cap. s\&h bias gen.& $C_{h,bias}$ & 307\,fF\\
			Unit cap. bias DAC & $C_u$ & 10\,fF\\
			Unit res. RF DAC & $R_u$ & 150\,$\Omega$\\
			\hline
			Bias memory activity & $\sigma_{biasmem}$ & $0.306$\\
			RF memory activity&$\sigma_{RFmem}$&$0.026$\\
			Managing comp. activity&$\sigma_{con}$&0.5\\
			Clock freq. bias  & $f_{clk, bias}$ & 2.22\,MHz\\
			Clock freq. RF  & $f_{clk, bias}$ & 600\,MHz\\

		\end{tabular}
	\end{ruledtabular}
\end{table}

Knowing the number of basic components, their sizes, and the technology data, we can estimate the area demand of the control circuit. With the added knowledge of the dynamic behaviour of the different circuit elements and of the system specifications, we can estimate the power consumption. The power consumption of the analogue parts of the DACs is approximately given by $P_{R}=V_{range}^2/R_{in}$ and $P_{C}=0.5\cdot f_{refresh}C_{in}V_{range}^2$, where 
$V_{range}$ is the typical voltage supplied to the input and $R_{in}$ and $C_{in}$ are the input resistance and capacitance of the DACs respectively. The supplied voltages are equal to the signal amplitudes $V_{range,bias}$ and $V_{range,RF} $ as given in Tab.\,\ref{tab:spec}. The power consumption of the digital parts and of the switches is modeled as
\begin{equation}\label{eq:Pswitch}
P_{dig}=P_{switch}=\sigma \cdot f_{clk}\cdot V_{dd}^2\cdot C_{gate},
\end{equation}
where $C_{gate}$ is the total capacitance of the considered circuit or switch, $f_{clock}$ the clock frequency, $V_{dd}$ the applied potential, and $\sigma$ is the activity of the circuit. For a simple switch $\sigma=0.5$, which is also the maximum in common clocked, edge triggered logic.  In the following we will conservatively assume $\sigma=0.5$ for most digital circuits, except for the memory parts where $\sigma$ can be derived from the circuit architecture with little effort (see Tab. III).
 Determining the precise activity of the other circuits would require extensive statistical simulation or testing. In Eq.\,\ref{eq:Pswitch} we neglect the short-circuit power consumption as well as the leakage power \cite{Waser2012}, since in the considered technology they are small compared to the switching power $P_{switch}$, especially at cryogenic temperatures \cite{RoyMukhopadhyayMahmoodi-Meimand2003, Crowder2005, Assenmacher2003,ButzenRibas2006,JunDiestelhorstBelliniEtAl2006, GhibaudoBalestra1997}.

\begin{figure}
\includegraphics[width=\columnwidth]{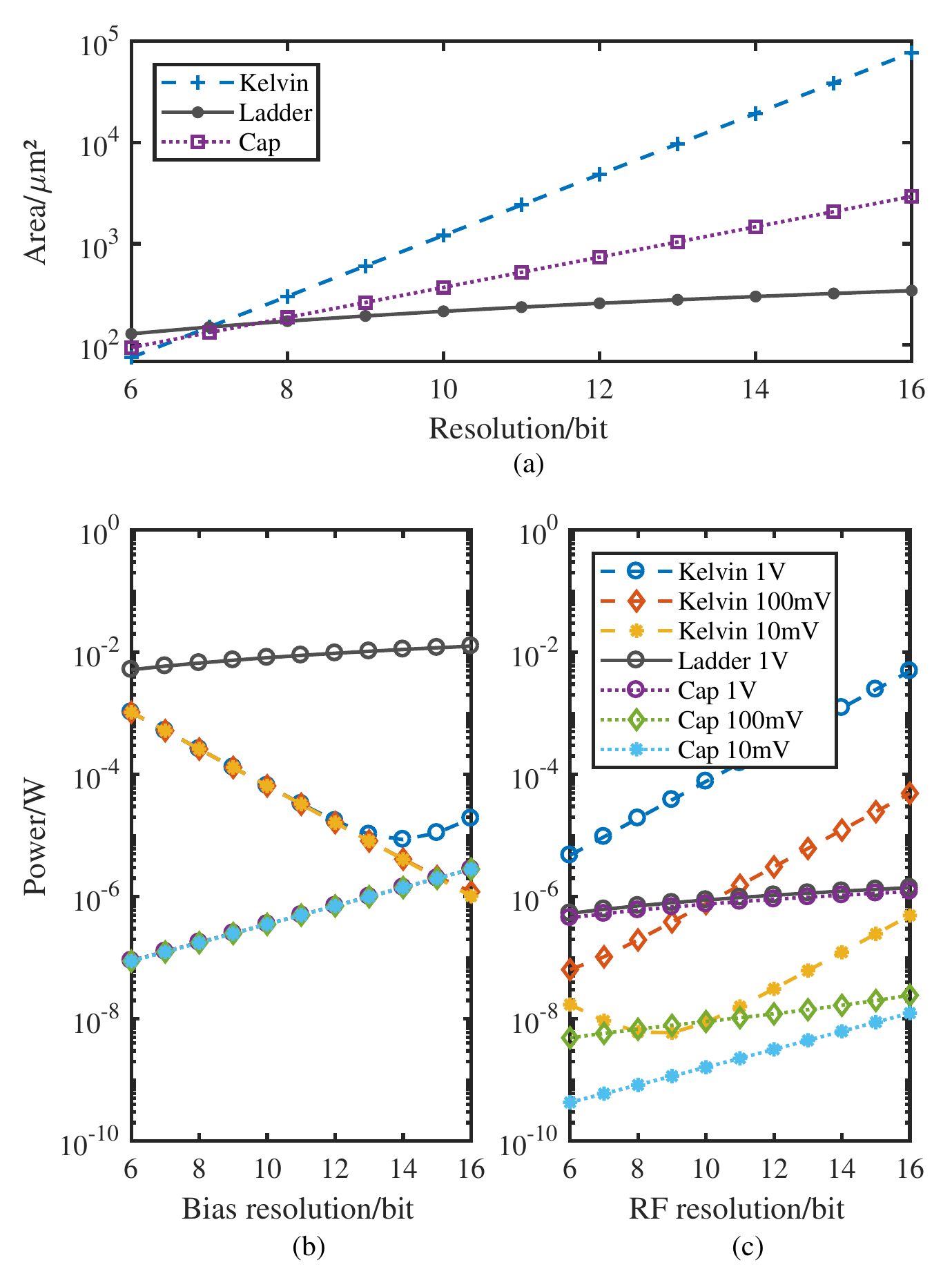}
		\caption{ Comparison of DACs with component dimensions as in Sec.\,\ref{subsec:bias} and Sec.\,\ref{subsec:rf} but with $R_{u,Ladder}=150\,\Omega$ a) Area consumption of the different DAC architectures for different resolutions, independent of operating conditions b) Power consumption of the DAC architecture, under the operating condition of the bias generation unit ($V_{range}=1\,$V and $f_{refresh}=1.11\,$MHz).  c) Same as in b), but now considering the operating conditions of the RF generation unit ($V_{range}=4\,$mV and $f_{refresh}=300\,$MHz). \label{pic:powerdac}}
\end{figure}

We first analyze the area and the power consumption of the three DAC architectures presented in Fig.\,\ref{pic:dac}. Current experiments employ a resolution of 12 and 10\,bit for bias and RF-signals, respectively. However, since these specifications are likely to change in the future, we plot our results as a function of resolution $n$. The area required by the three types of DAC is represented in Fig. 7a, and clearly point at the Ladder DAC as the most area-saving architecture, even with a unit resistance significantly larger than the one of the Kelvin DAC.  The power consumption of the three DACs is plotted in Fig. 7b and c, distinguishing between the typical operation conditions of the bias generation unit ($V_{range}=1\,$V and $f_{refresh}=1.11\,$MHz) and of the RF generation unit ($V_{range}=4\,$mV and $f_{refresh}=300\,$MHz). The power consumption is also plotted for several different supply voltages for the Cap and Kelvin DACs, as this has a major quantifiable impact on their dissipation and lower $V_{dd}$ may become available for future technologies optimized for cryogenic operation \cite{DATE_2017}. Vice versa, the power required by the Ladder DACs has little variation with $V_{dd}$ and is only plotted for $V_{dd}=1$\,V, which is the supply voltage in the current technology.

 In Fig.\,\ref{pic:powerdac}b,c the Kelvin DAC shows at times a counter intuitive behaviour of the power consumption over the resolution. The reason is that while $n$ is rising also $R_{in}$ increases, leading to a lower power consumption with a fixed $R_{u}$. For higher resolution however the power consumption of the switches inside the DAC dominates the total power consumption, thus leading to an increase.

The Cap DAC turns out to be the most power saving architecture, achieving a power consumption of around $10^{-7}\,$W for a DAC in the bias generation condition and a consumption in the range of $10^{-9}\,$W for the RF generation condition. This is in part thanks to the fact that this type of DAC has only dynamic power consumption, in contrast to resistor-based designs with static power consumption. The somewhat higher area consumption of the Cap DAC is compensated for by the significantly lower power consumption in case of the bias operation condition. For the RF operation condition the potential for a significant power reduction for smaller $V_{dd}$, in contrast to the Ladder DAC, is the criteria for the choice of the Cap DAC. In all the following estimates, we will therefore assume the DAC of the bias generation unit  and the RF generation unit to be a Cap DAC.

\begin{figure}
		\includegraphics[width=\columnwidth]{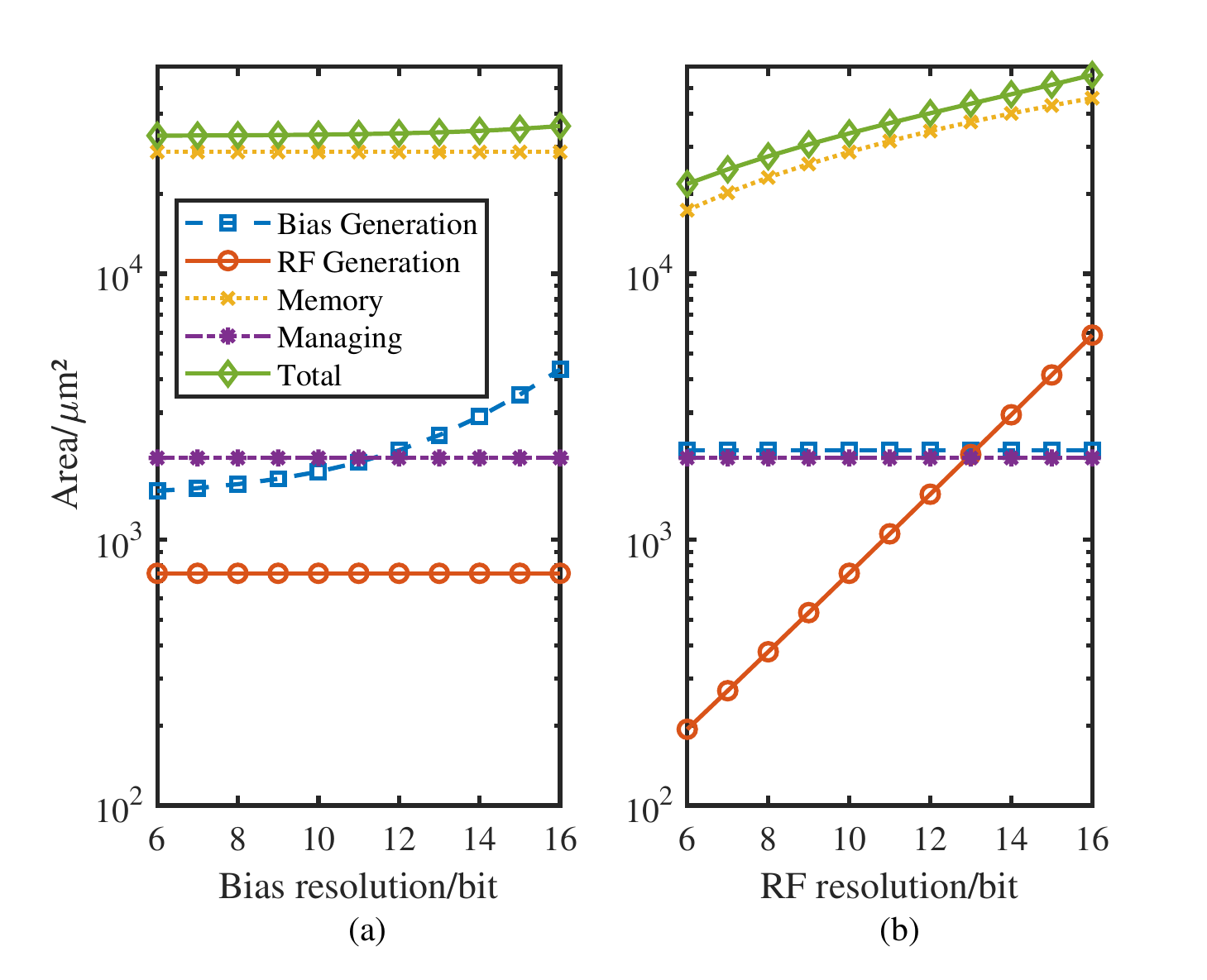}	
	\caption{Area consumption of the different components of the control electronics (labeled as in Fig.\,\ref{pic:complete}) for different values of the resolution of the bias signal (a) and of the RF signal (b). In panel a) the RF resolution is assumed to be 10\,bits, while in panel b) the bias resolution is fixed to 12\,bits.  
	\label{pic:area}}
\end{figure}

We now turn discussing the behavior of the whole control electronics. The total area consumption is shown in Fig.\,\ref{pic:area} as a function of the resolutions of the bias and the RF signals. For the typical resolutions used in current experiments (bias resolution = 12\,bits and RF resolution = 10\,bits) the control electronic would occupy an area of roughly $3.3\cdot10^4\,\mu \text{m}^2$, with the largest share taken up by the memory which is the only part with a significant dependence on the resolution.

\begin{figure}
		\includegraphics[width=\columnwidth]{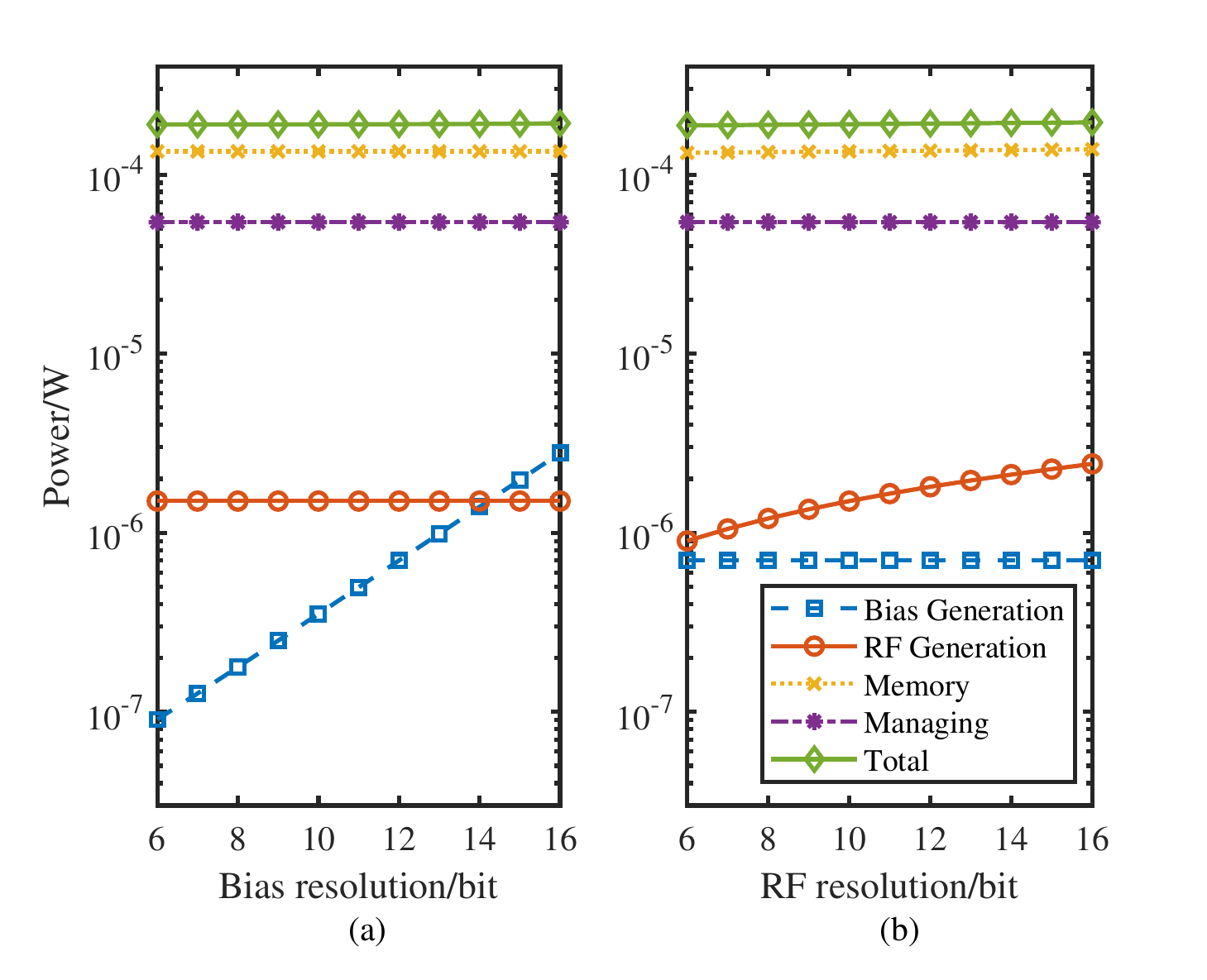}
	\caption{Power consumption of the different components of the control electronics (labeled as in Fig.\,\ref{pic:complete}) for different values of the resolution of the bias signal (a) and of the RF signal (b) using $V_{dd}=1\,$V. In panel a) the RF resolution is assumed to be 10\,bits, while in panel b) the bias resolution is fixed to 12\,bits. Both panels refer to the power consumption during qubit operation. In contrast to Fig.\,\ref{pic:powerdac}, the bias generation power includes the DAC and the s\&h, and the RF generation power includes the power of two DACs }\label{pic:power-cons}
\end{figure}

The estimated power consumption of the whole control electronics in the so--called ``qubit operation regime'', when the qubit is manipulated by RF pulses to perform various operations (e.g. logical gates), is shown in Fig.\,\ref{pic:power-cons}. In this regime, the clock frequency of the digital circuit elements $f_{clk}$ entering Eq.\,\ref{eq:Pswitch} is determined by $f_{refesh}$ of the associated analog parts of the DACs, with $f_{clk,bias}=f_{clk,biasmem}=f_{clk,biasgen}=2\cdot f_{refresh,bias}=2.22\,$MHz and $f_{clk,RF}=f_{clk,RFmem}=f_{clk,RFgen}=2\cdot f_{refresh,RF}=600$\,MHz. The factor 2 is included because logic gates are edge triggered (e.g. by the rising edge) and thus operate at half the frequency. The subunits of the managing component are running on different frequencies, as detailed in App.\,\ref{app:managing}. The total power consumption of the circuit with the current specifications ($V_{dd}=1$\,V,$ V_{range, bias}=1$\,V,$ V_{range, RF}= 4$\,mV, $f_{refresh,bias}= 1.11$\,MHz, $f_{refresh,RF}= 300$\,MHz)
is around $1.9\cdot10^{-4}$\,W, with almost no dependence on the resolution (see Fig.\,\ref{pic:power-cons}). More than 95\,\% of this is due to digital circuits, with the bias and the RF generation units accounting only for a few $\mu$W.

\begin{figure}
	\includegraphics[width=\columnwidth]{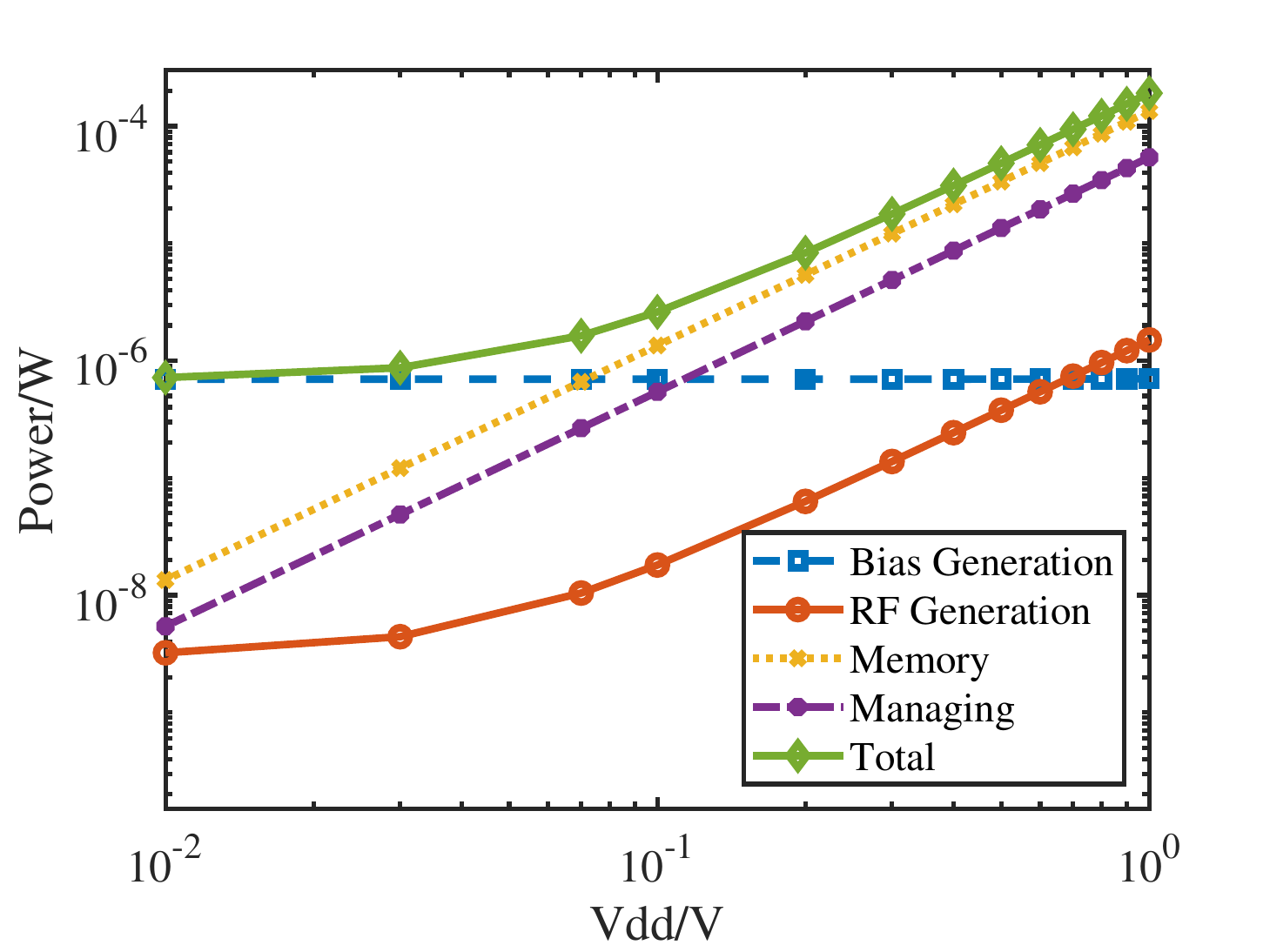}
		\caption{Power consumption of the different components of the control electronics (labeled as in Fig.\,\ref{pic:complete}) for different values of the digital supply voltage $V_{dd}$. The bias resolution is set to 12\,bit while the RF resolution is assumed to be 10\,bit and the figure again refers to the power consumption during qubit operation. For $V_{dd}<70$\,mV the bias generation unit dominates the total power consumption}\label{pic:power-vdd}
\end{figure}

Lowering the supply voltage as shown in Fig.\,\ref{pic:power-vdd} drastically reduces the percentage of the total power consumption the digital circuits and especially the memory and managing unit have. Below 70\,mV the bias generation part consumes the most power. 

\section{Discussion and outlook}
\label{sec:discussion}

\begin{table}
	\caption{Area and power consumption for different technology and architecture options included in the model: flip-flop memory (FF), SRAM memory (S), $V_{dd}=$digital supply voltage, Node=CMOS technology generation
		\label{tab:areapowerscale}}
	\begin{ruledtabular}
		\begin{tabular}{lcccc}
			Settings& &  &  & \\
			\hline 
			Node/nm & 65 & 65 &65 & 14 \\
			V$_{dd}$/V 	  & 1  &  1& 100m & 10m \\
			Architecture & FF & S &S &S\\
			\hline
			Area/$\mu$m$^2$&&&&\\
			\hline
			Bias Gen A. & $2.2 \cdot 10^3$ 		&$2.2 \cdot 10^3$ 		&$2.2 \cdot 10^3$ 		& $1.1 \cdot 10^1$ \\
			RF Gen A. & 	 $7.5 \cdot 10^2$ 		&$7.5 \cdot 10^2$ 		&$7.5 \cdot 10^2$ 		& $3.8 \cdot 10^0$  \\
			Memory A. &	 $2.9 \cdot 10^4$	 	&$2.6 \cdot 10^3$ 		&$2.6 \cdot 10^3$	 	& $2.1 \cdot 10^2$ \\
			Managing A. & $2.0 \cdot 10^3$ 		&$1.7 \cdot 10^3$ 		&$1.7 \cdot 10^3$	 	& $7.0 \cdot 10^1$ \\
			Total A. &	 $3.3 \cdot 10^4$ 		   &$7.2 \cdot 10^3$ 		&$7.2 \cdot 10^3$ 		& $3.0 \cdot 10^2$ \\
			\hline	
			Power/W&&&&\\
			\hline
			Bias Gen P. &$7.0 \cdot 10^{-7}$ 	&$7.0 \cdot 10^{-7}$ 	&$7.0 \cdot 10^{-7}$ 	& $7.0 \cdot 10^{-7}$ \\
			RF Gen P. & 	$1.5 \cdot 10^{-6}$ 	&$1.5 \cdot 10^{-6}$ 	&$1.8 \cdot 10^{-8}$ 	& $3.2 \cdot 10^{-9}$ \\
			Memory P. &	$1.3 \cdot 10^{-4}$ 	&$5.0 \cdot 10^{-5}$ 	&$5.0 \cdot 10^{-7}$ 	& $3.6 \cdot 10^{-9}$ \\
			Managing P. &$5.4 \cdot 10^{-5}$  	&$2.8 \cdot 10^{-5}$ 	&$2.8 \cdot 10^{-7}$ 	& $2.2 \cdot 10^{-9}$ \\
			Total P. &	$1.9 \cdot 10^{-4}$ 	&$8.1 \cdot 10^{-5}$ 	    &$1.5 \cdot 10^{-6}$ 	   & $7.0 \cdot 10^{-7}$ \\
		\end{tabular}
	\end{ruledtabular}
\end{table}

The estimates of area and power consumption presented in the previous section clearly point at some of the most relevant technological challenges that the vision of a quantum computer as illustrated in Fig.\,\ref{pic:levels} will have to face. 

According to our analysis, the control electronics for a single qubit will occupy an area
which is about four orders of magnitude larger than the current qubit footprint associated with the average interqubit spacing. The qubit distance allowing for entangling is up to a few microns for charge-coupled semiconductor qubits \cite{TrifunovicDialTrifEtAl2012} and less than 1$\,\mu$m for exchange coupled semiconductor qubits \cite{ZajacHazardMiEtAl2016}. Larger spacings are possible when using cavity coupling \cite{MiBenitoPutzEtAl2017,LandigKoskiScarlinoEtAl2018,SamkharadzeZhengKalhorEtAl2017} and for superconducting qubits. This discrepancy limits the possibility of placing the control electronics directly on top of each qubit in a scalable way with the specific design considered so far. The memory unit consumes the most area of all components, as depicted in Fig.\,\ref{pic:area}, which is also not greatly changed through a decrease in resolution of bias or RF signals. However, a significant area reduction is explored in Tab.\,\ref{tab:areapowerscale} by comparing two different memory types, the current flip-flop (F) and SRAM (S). 

SRAM is a much more area efficient, low-latency and non-volatile memory architecture, whose memory cells have a typical size of $0.5-0.7$\,$\mu\text{m}^2$ in a 65\,nm technology \cite{ArnaudBoeufSalvettiEtAl2003}, compared to the current $10$\,$\mu\text{m}^2$ of a flip-flop. However, SRAM functionality is based on sensitive analog device-properties, which are not yet well characterized at low temperatures. If SRAM proves reliable functionality at cryogenic temperatures, the area of the memory could be reduced by one order of magnitude (Tab.\,\ref{tab:areapowerscale}). 

Moving from the current 65\,nm technology to a smaller node like a 14\,nm technology will reduce the area of a logic transistor by a factor of 24 and the SRAM cell by a factor of 7 \cite{BorkarBohrJourdan2014,Intel10nm2017}. The capacitor and resistor densities are not directly affected by CMOS scaling, but the capacitive density could potentially be increased by a factor of more than 200 through the use of so called trench capacitors  \cite{GlobalFoundaries2018,Note3}.\footnotetext[3]{ \label{foot:3}. They have their capacitors placed vertically instead of horizontally and are available in new packaging solutions, where e.g. several circuit dies get bonded on top of each other before being packaged (3D packaging).}

In summary, exploiting the discussed and currently available technologies it should be possible to reduce the area consumption of the control electronics from $3.3\cdot 10^{4}\approx180 \times 180$\,$\mu\text{m}^2$ to $3.0\cdot 10^2\approx 20 \times 20 $\,$\mu\text{m}^2$ (Tab.\,\ref{tab:areapowerscale}). Some additional leeway for reducing the area discrepancy between qubit and electronics is offered by the not yet modeled stacking of dies, and by the current efforts of implementing long--range coupling of qubits, which would extend the footprint of the qubit. Literature suggests that a coupling range of 10\,$\mu$m might be feasible \cite{TrifunovicDialTrifEtAl2012,SerinaKloeffelLoss2017,ElmanBartlettDoherty2017}. Altogether, this indicates that with future research the area matching of qubit footprint and control electronics is in reach.

For the power consumption, our estimates indicate that with current technologies the control electronics would consume around 190\,$\mu$W. This has to be compared with the cooling power of the cryostats hosting the quantum computing systems. Current cryogen-free
dilution refrigerators supply up to a few mW at 200\,mK, less than 1\,W
at 1\,K, and a few W at 4\,K. The cryogenic qubit control systems
considered here would then be suitable for a 5 qubit demonstrator system (Tab.\,\ref{tab:qubitnumber}).

The already discussed options for the area optimization SRAM and the shrinking of the feature size present a complementary reduction in power consumption by approximately two third. 
In addition, a considerable reduction of power could be achieved by reducing the operating voltage of the digital elements, see Tab.\,\ref{tab:areapowerscale}. Reducing $V_{dd}$ from 1\,V to 100\,mV, which may be possible with reasonable process modifications to shift the transistor threshold voltage, would reduce the dissipation due to digital circuitry by a factor 100. 
This would in turn allow the operation of more than three hundred qubits. This number is already very inconvenient to achieve with standard dilution refrigerators and a control based on room temperature electronics. Reducing $V_{dd}$ to 10\,mV, which is physically possible but requires major reoptimization of transistors for cryogenic operation, the power consumption of the digital circuits could be reduced to the nW regime, while the total power consumption is at 700\,nW. In this case, the analog bias  generation unit would become the by far the major source of dissipation (see Tab.\,\ref{tab:areapowerscale} last column).

\begin{table}
	\caption{Number of controllable qubits for different quantifiable scenarios: $T_ {el}$ = operating temperature of electronics, $\dot Q$ = cooling power, No. Qubits = $\dot Q$ / dissipation, architecture: flip-flop memory (FF), SRAM memory (S). The analog component sizes and derived parameters are adjusted for 1.8\,K, if needed.
		\label{tab:qubitnumber}}
	\begin{ruledtabular}
		\begin{tabular}{cccccc}
			$  T_{el}$& $\dot Q$ & Architecture & Node &  $V_{dd}$ & No. Qubits \\
			\hline 
			200 mK  &  1 mW & FF & 65 nm & 1 V &  5 \\
			200 mK  &  1 mW & S & 14 nm & 100 mV &  328 \\
			200 mK  &  1 mW & S &  14 nm & 10 mV & $ 1.4 \cdot 10^3$\\
			1.8 K   &  1 W & S & 65 nm & 1 V & $1.3 \cdot 10^5$ \\
			1.8 K   &  10 W & S & 14 nm & 10 mV & $6.1 \cdot 10^8$ \\
		\end{tabular}
	\end{ruledtabular}
\end{table}

Another possible approach to relax the constraints on the power consumption is to thermally isolate qubits from the electronics while allowing high-density, microfabricated interconnects. Increasing the operating temperature of the control electronics would in fact considerably relax the constraints on the thermal budget and allow the operation of hundred thousand, or even billions of qubits, see Tab.\,\ref{tab:qubitnumber}. Using a Helium liquifier plant, it should be even possible to deliver at least 100\,W around 2\,K (as an extreme example, each cooling plant used at LHC provides a cooling power of 2.4\,kW at 1.9\,K). 

For a possible scale up to qubit numbers in the range of millions, with reasonable cooling effort, additional strategies are needed to bring down the power consumption of the bias generation. The digital parts of the unit consume negligible power at this point with $P_{bias}\approx P_C$, the power consumption of the periodically charged capacitors, leaving 
\begin{equation}
P_{bias}=\frac{f_{refresh}}{2}(C_{in,DAC}V_{range,bias}^2+C_{s\&h}\delta V_{bias}^2).
\end{equation}

Using Eq.\,\ref{eq:refresh} and the known parameter values the second summand can be neglected giving
\begin{equation}
P_{bias}\approx\frac{N_{bias}C_{in,DAC}}{2\cdot R_{off}\delta V_{bias}C_{s\&h}}V_{range}^3.
\end{equation}
For $R_{off}$ the conservative room temperature value 1\,T$\Omega$ has been used in the model taking no leakage reduction at cryogenic temperatures into account. With an anyhow necessary optimization of the technology for low supply voltages in addition to that, the leakage current and the resulting effective $R_{off}$, at least for selected transistors, can be expected to change by two orders of magnitude \cite{Incandela2017}.
The voltage range $V_{range}$ in which the qubit has to be tuned depends not only on the requirements of the qubit but also on the uniformity of the substrate. A reduction from 1\,V to 0.5\,V would already lead to a power reduction of  nearly one order of magnitude due to the cubic term. With an additional reduction of the resolution from 12\,bit to 8\,bit a reduction by a factor of 32 can be achieved.
With these two changes the power consumption of the bias generation is safely in the nW range even if other possibilities for power reductions with less impact, such as fewer gate electrodes and a further increase of $C_{s\&h}$ (with an area trade-off), remain unexplored. Other strategies to further reduce area and power consumption are subjects of current research, and include the possibility of sharing units (e.g. some subunits of the managing component or the memory) between different qubits, functional optimization

\section{Conclusion}
\label{sec:conclusion}

The work at hand represents a concrete feasibility analysis of a scalable integrated control system for spin qubits. We derived the systems specification from the requirements of a GaAs spin qubit and developed a dedicated electrical system using an exemplary 65\,nm CMOS process. The most significant properties regarding the scalability, the area and power consumption, were investigated in detail for the complete system as well as all subunits. In addition the estimations of area and power consumption were extrapolated to a current state of the art 14\,nm process. The results show that the power consumption presents a bigger challenge for scalability than the area, which is at $3.2 \cdot 10^2 \mu$m$^2$. For a possible reduction of the power consumption the effect of reducing the digital supply voltage $V_{dd}$ from currently 1\,V to 10\,mV was detailed. This lowers the expected power consumption of all subunits, except for the bias generation, into the nW regime. However, our assumptions have been conservative in general, and with taking a slightly more optimistic estimation of the transistor leakage current at cryogenic temperatures also the bias generation power consumption could be reduced to the nW range, as well. 

With the 2019 to be available 12\,nm node of the fully-depleted silicon-on-insulator (FD-SOI) 22\,nm CMOS technology\footnote{Global foundries product brief 22FDX\label{foot:22nm}} ($V_{dd}=400$\,mV) expected to run on $V_{dd}=200$\,mV, a tool for a first scale up is in reach. With this technology, and the currently available 1\,mW of cooling power at 200\,mK, more than 300 qubits are anticipated to be controllable through the system this work proposes. This shows the near future advantage of dedicated integrated electronics since it would be quite cumbersome to control that amount of qubits with room temperature electronics. An increase in cooling power to 1\,W, for example through thermal isolation of qubit and electronics, would already expectably enable the control of

\appendix

\section{Managing Component} \label{app:managing}

\begin{figure}
	\includegraphics[scale=0.8]{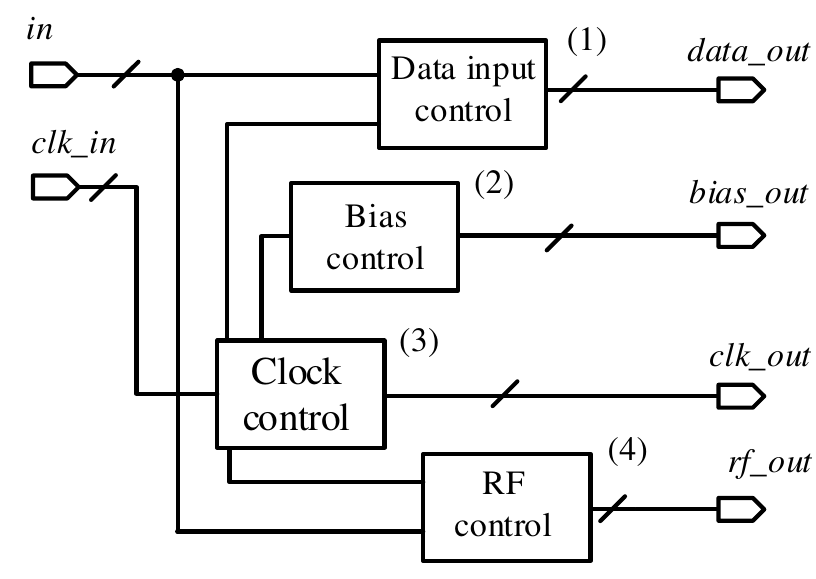}
	\caption{Block diagram of the managing component. 
	 \label{pic:steer}}
\end{figure}

The managing component consists of different subunits, each dedicated to a specific task, as depicted in Fig.\,\ref{pic:steer}. 

\begin{figure*}
	\includegraphics[scale=0.8]{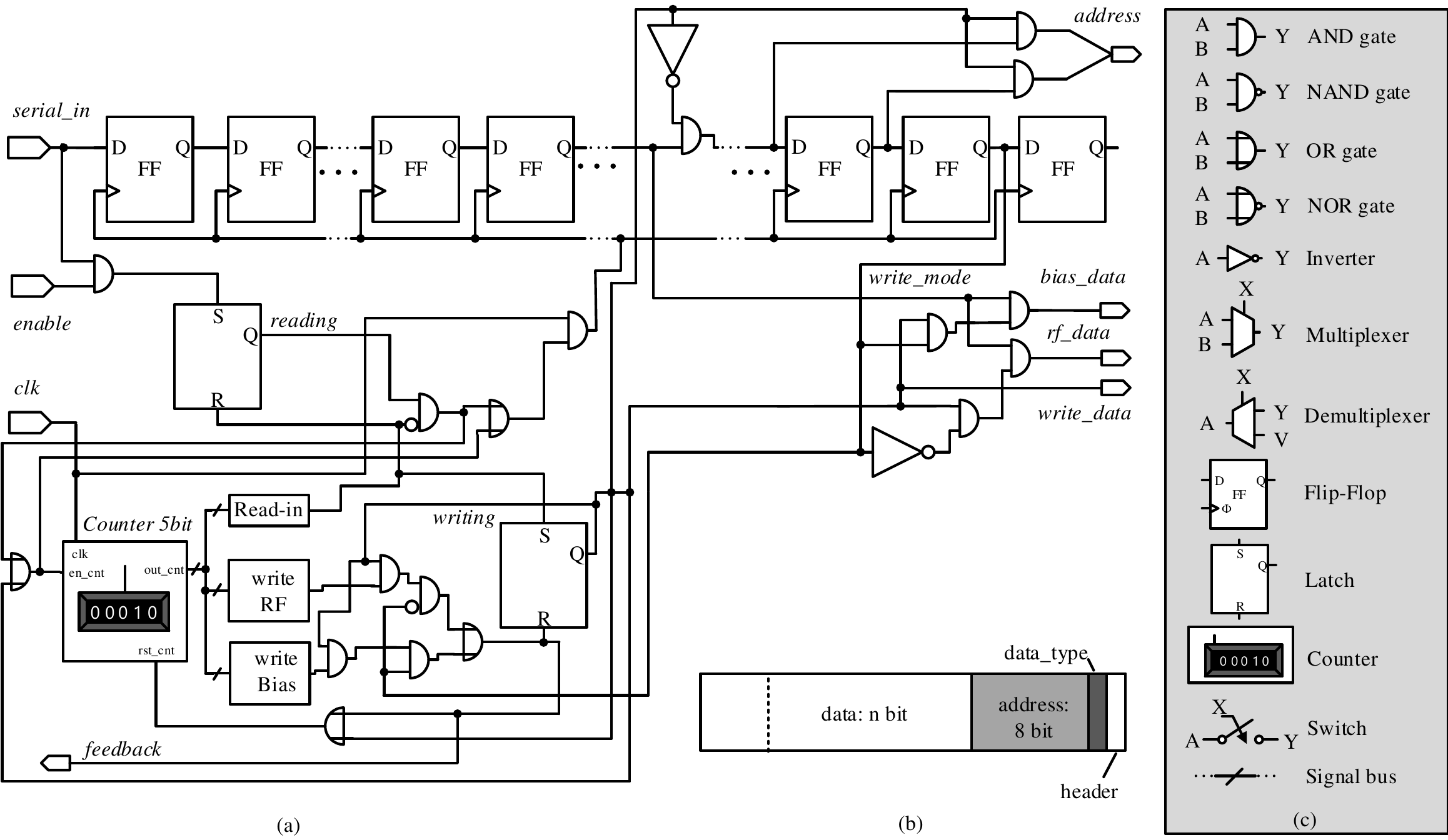}
	\caption{ a) Structure of the {\em data input control} unit, which receives voltage information to be stored in the memory, and distributes it with the corresponding address to the appropriate memory part b) Structure of the data word. c) Legend of the symbols used the block diagrams presented in the paper.
	\label{pic:managing-1}}
\end{figure*}
The data input control unit manages the reception of data and directs it to the correct memory part (Fig.\,\ref{pic:managing-1}). Data is received in data words of the form shown in Fig.\,\ref{pic:managing-1}b, with a first header bit of `1' and the second bit defining the data type, i.e.  bias or RF voltage information. The next part of the data word contains an 8\,bit long address to account for $256=2^8$ accessible registers in the RF memory. The data to be stored in the memory is located in last part of the data word, whose length is set by the required voltage resolution in the system.

Data words arrive serially at the {\em serial\_in} input, and get written into flip-flops connected as a shift register. When the transmission begins with the arrival of the header bit, a 5\,bit counter is started to time the reception of the data \cite{TietzeSchenk2008}. This happens through counting cycles of the clock signal {\em clock} which drives the data reception. The right number of passed clock cycles is determined by combinatorial logic, in combination with the counter which ends the data reception.

 At the end of the reception, the data address is available at the {\em address} bus connected to the respective {\em write\_select} inputs of the memory (see Fig.\,\ref{pic:memcomplete}). The write process to the memory is stared with a logical `1' at {\em write\_data} connected to the {\em enable} input at the memory.  The data itself is transported via the {\em bias\_data} and {\em rf\_data} outputs to the {\em data\_in} inputs of the memory.
 The timing of the writing process is determined again by the counter and some combinatorial logic. The end of the transmission and following writing process is signaled by an acknowledgment signal at the $feedback$ output. The data-input-control unit is only active when bias voltages or sequences are changed, and not during the ``qubit operation regime'' considered in Sec.\,\ref{sec:estimates}. For this reason, its power consumption is not included in the results of Fig.\,\ref{pic:power-cons} and Tab.\,\ref{tab:areapowerscale},\ref{tab:qubitnumber}. For a RF resolution of 10\,bit and a bias resolution of 12\,bit, the power consumption of the   data-input-control unit is 
$1.8\cdot 10^{-4}$\,W, with negligible dependence on the maximum system resolution.

\begin{figure*}
	\includegraphics[scale=0.8]{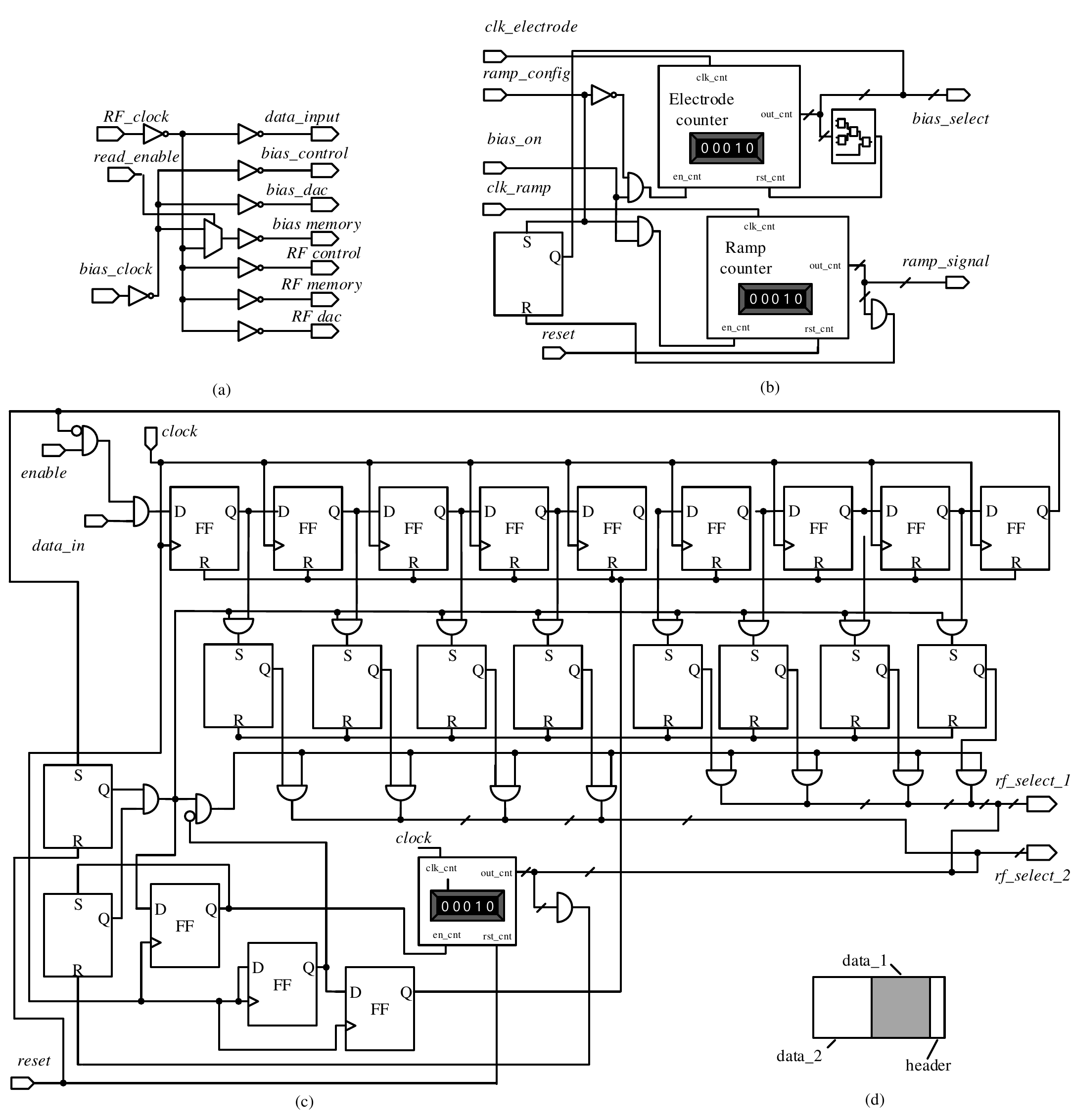}
	\caption{a)  Concept of the {\em clock control} unit, which distribute clock signals to the other subunits of the managing component. The inverters help improving edge steepness. b) Structure of the {\em bias control} unit, which steers the tuning of the qubit and the refresh of the voltages applied to the qubit electrodes. c) Concept of the {\em RF control} unit which, receives instructions on the sequences to apply for qubit operations. d) Structure of the RF data word. \label{pic:managing-2}}
\end{figure*}

The clock control unit receives stable digital reference--signals of different frequencies, and routes them to the other subunits. Its structure is shown in Fig.\,\ref{pic:managing-2}a. The use of in total two inverters (buffers) in each signal path enhances edge steepness, thus improving signal quality. 
The global clock signals, used also for general timing, are produced in a different stage of the quantum-computer architecture at higher temperatures. For minimization of power consumption, the digital control-units run at the minimal frequency  dictated by the corresponding analog part. 
Following this principle, we link the clock frequency of the bias control unit to the clock of the bias generation unit, with $f_{clk,bias}=2.22$\,MHz and the clock frequency of the RF control unit to the frequency of the RF generation part,  with $f_{clk,RF}=600$\,MHz. The frequency of the data--input--control unit can in principle be chosen arbitrarily but, to minimize the number of clock frequencies, we set it equal to $f_{clk,RF}$. Taking $f_{clk,bias}$ instead of $f_{clk,RF}$ is also a possible choice, but it slows down the data read-in by two orders of magnitude. The reduced power consumption through the lower frequency is no major advantage as long as possible temperature increase is lowered again before qubit operation.

The bias control unit has the structure shown Fig.\,\ref{pic:managing-2}b and manages the generation of bias signals through preparing the digital signal inputs of the bias generation unit. It either reads values from the bias memory  (see Fig.\,\ref{pic:memcomplete}a during gate operation or provides voltages values itself during dot tuning.

If {\em ramp\_config} is logical '0', the bias control unit is in the ``refresh mode'', and the voltages on all qubit electrodes get periodically refreshed during qubit operation.
In this mode the {\em bias\_select} output of the bias-control unit is connected not only to the  {\em read\_select} input of the bias memory, but also to the {\em electrode\_select} input of the bias-generation unit. This ensures that the voltage level stored in a certain memory register is available at the {\em dig\_in} (Fig.\,\ref{pic:biasgen}) input of the bias generation unit and the right voltage is applied to the right qubit electrode. The periodicity is achieved with an automatic reset of the counter, which counts through all electrodes. 

 If {\em ramp\_config} is logical '1', the unit is in the ``tuning mode'', which allows applying a voltage ramp to one of the electrodes forming the double dot -- a procedure which is necessary for the initial tuning-up of the qubit in the right operation regime \cite{Botzem2018}. In this case, the output $ramp$\_$signal$ is connected to the input of the bias generation unit, and the ramp counter generates a digital stepwise ramp, which is converted into an analog voltage ramp by the bias generation unit.  In the mean while, the {\em electrode\_select} input of the bias generation unit is connected to the {\em data\_out} of the memory, and determines to which electrode the voltage ramp is applied. The address of the electrode is stored in one bias memory register and the read availability at {\em data\_out} is activated through the {\em bias\_select} output. The electrode counter is deactivated in this mode.

Finally, the structure of the  RF control unit is shown in Fig.\,\ref{pic:managing-2}c. This unit is responsible for applying gates and their associated pulses to the qubit while the succession of gates is determined by the higher levels of the quantum computer. Since the voltage values of all $N_{pulses}=16$ sequences with their length of samples $l_{pulses}=16$ (see Sec.\,\ref{subsec:spec}, Tab.\,\ref{tab:spec}) are stored in the RF memory (Fig.\,\ref{pic:memcomplete}) only the information on which sequences to be applied needs to be communicated. The RF control unit serially receives data words in the form of Fig.\,\ref{pic:managing-2}d containing two sets of data next to a header. Each data set contains two 4\,bit sequence identifiers (IDs) with $2^4=16$ unique IDs from '0000' to '1111' each for one RF electrode. The reception of the data works as with the data input control. When the sequences applied before are finished ($end\_sequ=$'1') the values from the flip-flops are transferred to the array of latches leaving the flip-flops free for a new data input immediately. At the same time the complete sample addresses are constructed one after one from the IDs and the sample counter indicating the number of the sample in the current sequence. The addresses are supplied to the {\em read\_select1,2} memory inputs via the {\em rf\_select1,2} outputs with the memory outputs in turn {\em data\_out1,2} connected to the {\em dig\_in} inputs of the RF generation unit supplying the digital voltage values to be converted to analog voltages.

\begin{acknowledgments}

We thank F. Haupt for  helpful  input  on  the  manuscript. We acknowledge support by the Impulse and Networking Fund of the Helmholtz association.

\end{acknowledgments}

\bibliography{Lit_clean}

\end{document}